\documentclass[nofootinbib,prd]{revtex4}%
\usepackage{amsmath}
\usepackage{amsfonts}
\usepackage{amssymb}
\usepackage{graphicx}%
\begin{document}
\title{The induced Cosmological Constant as a tool for exploring geometries}
\author{Remo Garattini}
\email{Remo.Garattini@unibg.it}
\affiliation{Universit\`{a} degli Studi di Bergamo, Facolt\`{a} di Ingegneria, Viale
Marconi 5, 24044 Dalmine (Bergamo) ITALY.}
\affiliation{INFN - sezione di Milano, Via Celoria 16, Milan, Italy.}

\begin{abstract}
The cosmological constant induced by quantum fluctuation of the graviton on a
given background is considered as a tool for building a spectrum of different
geometries. In particular, we apply the method to the Schwarzschild background
with positive and negative mass parameter. In this way, we put on the same
level of comparison the related naked singularity $\left(  -M\right)  $ and
the positive mass wormhole. We use the Wheeler-De Witt equation as a basic
equation to perform such an analysis regarded as a Sturm-Liouville problem .
The cosmological constant is considered as the associated eigenvalue. The used
method to study such a problem is a variational approach with Gaussian trial
wave functionals. We approximate the equation to one loop in a Schwarzschild
background. A zeta function regularization is involved to handle with
divergences. The regularization is closely related to the subtraction
procedure appearing in the computation of Casimir energy in a curved
background. A renormalization procedure is introduced to remove the infinities
together with a renormalization group equation.

\end{abstract}
\maketitle

\section{Introduction}

In 1969, Penrose suggested that there might be a sort of \textquotedblleft
cosmic censor\textquotedblright\ that forbids naked singularities from
forming\cite{Penrose}, namely singularities that are visible to distant
observers. Although there is no proof of such conjecture, naked singularities
and the cosmic censorship are still a source of interest. A simple and
particularly interesting example of naked singularities is the negative mass
Schwarzschild spacetime. This is simply obtained by the Schwarzschild solution%
\begin{equation}
ds^{2}=-\left(  1-\frac{2MG}{r}\right)  dt^{2}+\left(  1-\frac{2MG}{r}\right)
^{-1}dr^{2}+r^{2}\left(  d\theta^{2}+\sin^{2}\theta d\phi^{2}\right)  ,
\end{equation}
replacing $M$ with $-M$. This simple substitution gives rise to a naked
singularity not protected by a horizon. An immediate consequence of a negative
Schwarzschild mass is that if one were to place two bodies initially at rest,
one with a negative mass and the other with a positive mass, both will
accelerate in the same direction going from the negative mass to the positive
one. Furthermore, if the two masses are of the same magnitude, they will
uniformly accelerate forever.%
\begin{figure}
[ptb]
\begin{center}
\includegraphics[
height=2.1439in,
width=1.6656in
]%
{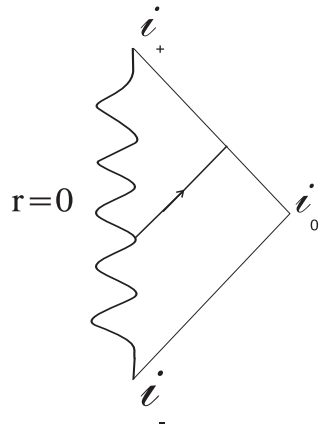}%
\caption{Carter Penrose diagram representing a naked singularity.}%
\end{center}
\end{figure}
This feature leads to the problem of stability of such a geometry discussed by
Gibbons, Hartnoll and Ishibashi\cite{GHI} and Gleiser and Dotti\cite{GD}.
However, even if the stability issue is an open debate which seems to incline
more to the unstable behavior than to the stable one\cite{GD1}, in this paper
we wish to study the relation between the Schwarzschild solution for positive
and negative masses with the induced cosmological constant. A cosmological
constant can be considered \textquotedblleft\textit{induced}\textquotedblright%
\ when it appears as a consequence of quantum fluctuations. Since, apparently
the Schwarzschild solution, naked singularities and the induced cosmological
constant appear to be disconnected, it urges to establish a point of contact.
We claim that such a link is in the Wheeler-De Witt equation (WDW)\cite{De
Witt}. This equation can be simply obtained starting by the Einstein field
equations without matter fields in four dimensions%
\begin{equation}
G_{\mu\nu}+\Lambda_{c}g_{\mu\nu}=0,
\end{equation}
where $G_{\mu\nu}$ is the Einstein tensor and $\Lambda_{c}$ is the
cosmological constant. By introducing a time-like unit vector $u^{\mu}$, we
get%
\begin{equation}
G_{\mu\nu}u^{\mu}u^{\mu}=\Lambda_{c}.\label{EoM}%
\end{equation}
This is simply the Hamiltonian constraint written in terms of equation of
motion. Indeed, if we multiply by $\sqrt{g}/\left(  2\kappa\right)  $
Eq.$\left(  \ref{EoM}\right)  $, we obtain $\left(  \kappa=8\pi G\right)  $
\begin{equation}
\frac{\sqrt{g}}{2\kappa}G_{\mu\nu}u^{\mu}u^{\mu}=\frac{\sqrt{g}}{2\kappa
}R+\frac{2\kappa}{\sqrt{g}}\left(  \frac{\pi^{2}}{2}-\pi^{\mu\nu}\pi_{\mu\nu
}\right)  =\frac{\sqrt{g}}{2\kappa}\Lambda_{c}.\label{EoM1}%
\end{equation}
Here $R$ is the scalar curvature in three dimensions and%
\begin{equation}
\frac{\sqrt{g}}{2\kappa}G_{\mu\nu}u^{\mu}u^{\mu}=-\mathcal{H}.
\end{equation}
If we multiply both sides of Eq.$\left(  \ref{EoM1}\right)  $ by $\Psi\left[
g_{ij}\right]  $, we can re-cast the equation in the following form%
\begin{equation}
\left[  \frac{2\kappa}{\sqrt{g}}G_{ijkl}\pi^{ij}\pi^{kl}-\frac{\sqrt{g}%
}{2\kappa}\left(  R-2\Lambda_{c}\right)  \right]  \Psi\left[  g_{ij}\right]
=0.\label{WDW}%
\end{equation}
This is known as the Wheeler-DeWitt equation with a cosmological term.
Eq.$\left(  \ref{WDW}\right)  $ together with%
\begin{equation}
-2\nabla_{i}\pi^{ij}\Psi\left[  g_{ij}\right]  =0,\label{diff}%
\end{equation}
describe the \textit{wave function of the universe}. The WDW equation
represents invariance under \textit{time} reparametrization in an operatorial
form, while Eq.$\left(  \ref{diff}\right)  $ represents invariance under
diffeomorphism. $G_{ijkl}$ is the \textit{supermetric} defined as%
\begin{equation}
G_{ijkl}=\frac{1}{2}(g_{ik}g_{jl}+g_{il}g_{jk}-g_{ij}g_{kl}).
\end{equation}
Note that the WDW equation can be cast into the form%
\begin{equation}
\left[  \frac{2\kappa}{\sqrt{g}}G_{ijkl}\pi^{ij}\pi^{kl}-\frac{\sqrt{g}%
}{2\kappa}R\right]  \Psi\left[  g_{ij}\right]  =-\sqrt{g}\frac{\Lambda_{c}%
}{\kappa}\Psi\left[  g_{ij}\right]  ,
\end{equation}
which formally looks like an eigenvalue equation. In this paper, we wish to
use the induced cosmological constant argument evaluated to one loop in the
different backgrounds as a tool to establish which kind of background induce
the larger cosmological constant or, in other words, the larger Zero Point
Energy (ZPE). In particular, we will compute the graviton ZPE propagating on
the Schwarzschild background which positive and negative mass (naked
singularity). This choice is dictated by considering that the Schwarzschild
solution represents the only non-trivial static spherical symmetric solution
of the Vacuum Einstein equations. Therefore, in this context the ZPE can be
attributed only to quantum fluctuations. An example of this method applied in
a completely different context without a cosmological term is in
Refs.\cite{DGL,RemoRN}, where the ZPE graviton contribution computed on
different metrics is compared. In practice, we desire to compute%
\begin{equation}
\Delta\Lambda_{c}=\Lambda_{c}^{S}-\Lambda_{c}^{N}\gtreqqless0,
\end{equation}
where $\Lambda_{c}^{S,N}$ are the induced cosmological constant computed in
the different backgrounds. Moreover, the Schwarzschild solution for both
masses, namely $\pm M$ is asymptotically flat. Therefore we are comparing
backgrounds with the same asymptotically behavior. Nevertheless, in
Eq.$\left(  \ref{WDW}\right)  $, surface terms never come into play because
$\mathcal{H}$ as well $\Lambda_{c}/\kappa$ are energy densities and surface
terms are related to the energy (e.g. ADM mass) and not to the energy density.
We want to point up that we are neither discussing the problem of forming the
naked singularity nor a transition during a gravitational collapse, rather the
singularity is considered already existing. The semi-classical procedure
followed in this work relies heavily on the formalism outlined in
Ref.\cite{Remo}, where the graviton one-loop contribution in a Schwarzschild
background was computed, through a variational approach with Gaussian trial
wave functionals. A zeta function regularization is used to deal with the
divergences, and a renormalization procedure is introduced, where the finite
one loop is considered as a self-consistent source for traversable wormholes.
Rather than reproducing the formalism, we shall refer the reader to [16] for
details, when necessary. The rest of the paper is structured as follows, in
section \ref{p1}, we show how to apply the variational approach to the
Wheeler-De Witt equation and we give some of the basic rules to solve such an
equation approximated to second order in metric perturbation, in section
\ref{p2}, we analyze the spin-2 operator or the operator acting on transverse
traceless tensors specified for the Schwarzschild metric with $\pm M$, in
section \ref{p3} we use the zeta function to regularize the divergences coming
from the evaluation of the ZPE for TT tensors and we write the renormalization
group equation. We summarize and conclude in section \ref{p4}.

\section{The cosmological constant as an eigenvalue of the Wheeler De Witt
Equation}

\label{p1}In this section we shall consider the formalism outlined in detail
in Ref.\cite{Remo}, where the graviton

one-loop contribution in a Schwarzschild background is used. We refer the
reader to Ref.\cite{Remo} for details. The WDW equation $\left(
\ref{WDW}\right)  $, written as an eigenvalue equation, can be cast into the
form%
\begin{equation}
\hat{\Lambda}_{\Sigma}\Psi\left[  g_{ij}\right]  =-\frac{\Lambda_{c}}{\kappa
}\Psi\left[  g_{ij}\right]  , \label{WDW1}%
\end{equation}
where
\begin{equation}
\hat{\Lambda}_{\Sigma}=\frac{2\kappa}{\sqrt{g}}G_{ijkl}\pi^{ij}\pi^{kl}%
-\frac{\sqrt{g}}{2\kappa}R.
\end{equation}
We, now multiply Eq.$\left(  \ref{WDW1}\right)  $ by $\Psi^{\ast}\left[
g_{ij}\right]  $ and we functionally integrate over the three spatial metric
$g_{ij}$, then after an integration over the hypersurface $\Sigma$, one can
formally re-write the WDW equation as%
\begin{equation}
\frac{1}{V}\frac{\int\mathcal{D}\left[  g_{ij}\right]  \Psi^{\ast}\left[
g_{ij}\right]  \int_{\Sigma}d^{3}x\hat{\Lambda}_{\Sigma}\Psi\left[
g_{ij}\right]  }{\int\mathcal{D}\left[  g_{ij}\right]  \Psi^{\ast}\left[
g_{ij}\right]  \Psi\left[  g_{ij}\right]  }=\frac{1}{V}\frac{\left\langle
\Psi\left\vert \int_{\Sigma}d^{3}x\hat{\Lambda}_{\Sigma}\right\vert
\Psi\right\rangle }{\left\langle \Psi|\Psi\right\rangle }=-\frac{\Lambda_{c}%
}{\kappa}. \label{WDW2}%
\end{equation}
The formal eigenvalue equation is a simple manipulation of Eq.$\left(
\ref{WDW}\right)  $. However, we gain more information if we consider a
separation of the spatial part of the metric into a background term, $\bar
{g}_{ij}$, and a perturbation, $h_{ij}$,%
\begin{equation}
g_{ij}=\bar{g}_{ij}+h_{ij}.
\end{equation}
The perturbation can be decomposed in a canonical way to
give\cite{BergerEbin,York,MazurMottola,Vassilevich}%
\begin{equation}
h_{ij}=\frac{1}{3}\left(  h+2\nabla\cdot\xi\right)  g_{ij}+\left(
L\xi\right)  _{ij}+h_{ij}^{\bot} \label{p21a}%
\end{equation}
where the operator $L$ maps $\xi_{i}$ into symmetric tracefree tensors%

\begin{equation}
\left(  L\xi\right)  _{ij}=\nabla_{i}\xi_{j}+\nabla_{j}\xi_{i}-\frac{2}%
{3}g_{ij}\left(  \nabla\cdot\xi\right)
\end{equation}
and%
\begin{equation}
g^{ij}h_{ij}^{\bot}=0,\qquad\nabla^{i}h_{ij}^{\bot}=0.
\end{equation}
It is immediate to recognize that the trace element%
\begin{equation}
\sigma=h+2\left(  \nabla\cdot\xi\right)  \label{sigma}%
\end{equation}
is gauge invariant. We write the trial wave functional as%
\begin{equation}
\Psi\left[  h_{ij}\left(  \overrightarrow{x}\right)  \right]  =\mathcal{N}%
\Psi\left[  h_{ij}^{\bot}\left(  \overrightarrow{x}\right)  \right]
\Psi\left[  h_{ij}^{\Vert}\left(  \overrightarrow{x}\right)  \right]
\Psi\left[  \sigma\left(  \overrightarrow{x}\right)  \right]  ,\label{twf}%
\end{equation}
where
\begin{equation}%
\begin{array}
[c]{c}%
\Psi\left[  h_{ij}^{\bot}\left(  \overrightarrow{x}\right)  \right]
=\exp\left\{  -\frac{1}{4}\left\langle hK^{-1}h\right\rangle _{x,y}^{\bot
}\right\}  \\
\\
\Psi\left[  h_{ij}^{\Vert}\left(  \overrightarrow{x}\right)  \right]
=\exp\left\{  -\frac{1}{4}\left\langle \left(  L\xi\right)  K^{-1}\left(
L\xi\right)  \right\rangle _{x,y}^{\Vert}\right\}  \\
\\
\Psi\left[  \sigma\left(  \overrightarrow{x}\right)  \right]  =\exp\left\{
-\frac{1}{4}\left\langle \sigma K^{-1}\sigma\right\rangle _{x,y}%
^{Trace}\right\}
\end{array}
.
\end{equation}
The symbol \textquotedblleft$\perp$\textquotedblright\ denotes the
transverse-traceless tensor (TT) (spin 2) of the perturbation, while the
symbol \textquotedblleft$\Vert$\textquotedblright\ denotes the longitudinal
part (spin 1) of the perturbation. Finally, the symbol \textquotedblleft%
$trace$\textquotedblright\ denotes the scalar part of the perturbation.
$\mathcal{N}$ is a normalization factor, $\left\langle \cdot,\cdot
\right\rangle _{x,y}$ denotes space integration and $K^{-1}$ is the inverse
\textquotedblleft\textit{propagator}\textquotedblright. We will fix our
attention to the TT tensor sector of the perturbation representing the
graviton and the scalar sector. Therefore, representation $\left(
\ref{twf}\right)  $ reduces to%
\begin{equation}
\Psi\left[  h_{ij}\left(  \overrightarrow{x}\right)  \right]  =\mathcal{N}%
\exp\left\{  -\frac{1}{4}\left\langle hK^{-1}h\right\rangle _{x,y}^{\bot
}\right\}  \exp\left\{  -\frac{1}{4}\left\langle \sigma K^{-1}\sigma
\right\rangle _{x,y}^{Trace}\right\}  .\label{tt}%
\end{equation}
Actually there is no reason to neglect longitudinal perturbations. However,
following the analysis of Mazur and Mottola of Ref.\cite{MazurMottola} on the
perturbation decomposition, we can discover that the relevant components can
be restricted to the TT modes and to the trace modes. Moreover, for certain
backgrounds TT tensors can be a source of instability as shown in
Refs.\cite{Instability}. Even the trace part can be regarded as a source of
instability. Indeed this is usually termed \textit{conformal }instability. The
appearance of an instability on the TT modes is known as non conformal
instability. This means that does not exist a gauge choice that can eliminate
negative modes. Since the wave functional $\left(  \ref{tt}\right)  $
separates the degrees of freedom, we assume that%
\begin{equation}
-\frac{\Lambda_{c}}{\kappa}=-\frac{\Lambda_{c}^{\bot}}{\kappa}-\frac
{\Lambda_{c}^{trace}}{\kappa},
\end{equation}
then Eq.$\left(  \ref{WDW2}\right)  $ becomes%
\begin{equation}
\frac{1}{V}\frac{\left\langle \Psi\left\vert \hat{\Lambda}_{\Sigma}^{\bot
}\right\vert \Psi\right\rangle }{\left\langle \Psi|\Psi\right\rangle }%
=-\frac{\Lambda_{c}^{\bot}}{\kappa}\label{LambdaTT}%
\end{equation}
and%
\begin{equation}
\frac{1}{V}\frac{\left\langle \Psi\left\vert \hat{\Lambda}_{\Sigma}%
^{trace}\right\vert \Psi\right\rangle }{\left\langle \Psi|\Psi\right\rangle
}=-\frac{\Lambda_{c}^{trace}}{\kappa}\label{svareq}%
\end{equation}

\section{The transverse traceless (TT) spin 2 operator and the W.K.B.
approximation}

\label{p2}Extracting the TT tensor contribution from Eq.$\left(
\ref{LambdaTT}\right)  $, we get to one loop%
\begin{equation}
\frac{\Lambda_{c}^{\bot}}{\kappa}\left(  \lambda_{i}\right)  =-\frac{1}{4}%
{\displaystyle\sum_{\tau}}
\left[  \sqrt{\omega_{1}^{2}\left(  \tau\right)  }+\sqrt{\omega_{2}^{2}\left(
\tau\right)  }\right]  . \label{lambda1loop}%
\end{equation}
The above expression makes sense only for $\omega_{i}^{2}\left(  \tau\right)
>0$. To further proceed, we need to compute $\omega_{i}^{2}\left(
\tau\right)  $ $\left(  i=1,2\right)  $. To this purpose we write the
background metric in the following way%
\begin{equation}
ds^{2}=-N^{2}\left(  r\right)  dt^{2}+\frac{dr^{2}}{1-\frac{b\left(  r\right)
}{r}}+r^{2}\left(  d\theta^{2}+\sin^{2}\theta d\phi^{2}\right)  ,
\label{metric}%
\end{equation}
with a generic $b\left(  r\right)  $, to keep the discussion on a general
ground, when possible. $N\left(  r\right)  $ is the \textquotedblleft%
\textit{lapse function}\textquotedblright\ playing the role of the
\textquotedblleft\textit{redshift function}\textquotedblright, while $b\left(
r\right)  $ is termed \textquotedblleft\textit{shape function}%
\textquotedblright. The Spin-two operator for this metric is%
\begin{equation}
\left(  \triangle_{2}h^{TT}\right)  _{i}^{j}:=-\left(  \triangle_{T}%
h^{TT}\right)  _{i}^{j}+2\left(  Rh^{TT}\right)  _{i}^{j}, \label{spin2}%
\end{equation}
where the transverse-traceless (TT) tensor for the quantum fluctuation is
obtained with the help of Eq.$\left(  \ref{p21a}\right)  $. Thus%
\begin{equation}
-\left(  \triangle_{T}h^{TT}\right)  _{i}^{j}=-\triangle_{S}\left(
h^{TT}\right)  _{i}^{j}+\frac{6}{r^{2}}\left(  1-\frac{b\left(  r\right)  }%
{r}\right)  . \label{tlap}%
\end{equation}
$\triangle_{S}$ is the scalar curved Laplacian, whose form is%
\begin{equation}
\triangle_{S}=\left(  1-\frac{b\left(  r\right)  }{r}\right)  \frac{d^{2}%
}{dr^{2}}+\left(  \frac{4r-b^{\prime}\left(  r\right)  r-3b\left(  r\right)
}{2r^{2}}\right)  \frac{d}{dr}-\frac{L^{2}}{r^{2}} \label{slap}%
\end{equation}
and $R_{j\text{ }}^{a}$ is the mixed Ricci tensor whose components are:
\begin{equation}
R_{i}^{a}=\left\{  \frac{b^{\prime}\left(  r\right)  }{r^{2}}-\frac{b\left(
r\right)  }{r^{3}},\frac{b^{\prime}\left(  r\right)  }{2r^{2}}+\frac{b\left(
r\right)  }{2r^{3}},\frac{b^{\prime}\left(  r\right)  }{2r^{2}}+\frac{b\left(
r\right)  }{2r^{3}}\right\}  ,
\end{equation}
This implies that the scalar curvature is traceless. We are therefore led to
study the following eigenvalue equation
\begin{equation}
\left(  \triangle_{2}h^{TT}\right)  _{i}^{j}=\omega^{2}h_{j}^{i} \label{p31}%
\end{equation}
where $\omega^{2}$ is the eigenvalue of the corresponding equation. In doing
so, we follow Regge and Wheeler in analyzing the equation as modes of definite
frequency, angular momentum and parity\cite{Regge Wheeler}. In particular, our
choice for the three-dimensional gravitational perturbation is represented by
its even-parity form%
\begin{equation}
h_{ij}^{even}\left(  r,\vartheta,\phi\right)  =diag\left[  H\left(  r\right)
\left(  1-\frac{b\left(  r\right)  }{r}\right)  ^{-1},r^{2}K\left(  r\right)
,r^{2}\sin^{2}\vartheta K\left(  r\right)  \right]  Y_{lm}\left(
\vartheta,\phi\right)  . \label{pert}%
\end{equation}
For a generic value of the angular momentum $L$, representation $\left(
\ref{pert}\right)  $ joined to Eq.$\left(  \ref{tlap}\right)  $ lead to the
following system of PDE's%

\begin{equation}
\left\{
\begin{array}
[c]{c}%
\left(  -\triangle_{S}+\frac{6}{r^{2}}\left(  1-\frac{b\left(  r\right)  }%
{r}\right)  +2\left(  \frac{b^{\prime}\left(  r\right)  }{r^{2}}%
-\frac{b\left(  r\right)  }{r^{3}}\right)  \right)  H\left(  r\right)
=\omega_{1,l}^{2}H\left(  r\right) \\
\\
\left(  -\triangle_{S}+\frac{6}{r^{2}}\left(  1-\frac{b\left(  r\right)  }%
{r}\right)  +2\left(  \frac{b^{\prime}\left(  r\right)  }{2r^{2}}%
+\frac{b\left(  r\right)  }{2r^{3}}\right)  \right)  K\left(  r\right)
=\omega_{2,l}^{2}K\left(  r\right)
\end{array}
\right.  , \label{p33}%
\end{equation}
Defining reduced fields%

\begin{equation}
H\left(  r\right)  =\frac{f_{1}\left(  r\right)  }{r};\qquad K\left(
r\right)  =\frac{f_{2}\left(  r\right)  }{r},
\end{equation}
and passing to the proper geodesic coordinate%
\begin{equation}
dx=\pm\frac{dr}{\sqrt{1-\frac{b\left(  r\right)  }{r}}}, \label{throat}%
\end{equation}
the system $\left(  \ref{p33}\right)  $ becomes%

\begin{equation}
\left\{
\begin{array}
[c]{c}%
\left[  -\frac{d^{2}}{dx^{2}}+V_{1}\left(  r\right)  \right]  f_{1}\left(
x\right)  =\omega_{1,l}^{2}f_{1}\left(  x\right)  \\
\\
\left[  -\frac{d^{2}}{dx^{2}}+V_{2}\left(  r\right)  \right]  f_{2}\left(
x\right)  =\omega_{2,l}^{2}f_{2}\left(  x\right)
\end{array}
\right.  \label{p34}%
\end{equation}
with
\begin{equation}
\left\{
\begin{array}
[c]{c}%
V_{1}\left(  r\right)  =\frac{l\left(  l+1\right)  }{r^{2}}+U_{1}\left(
r\right)  \\
\\
V_{2}\left(  r\right)  =\frac{l\left(  l+1\right)  }{r^{2}}+U_{2}\left(
r\right)
\end{array}
\right.  ,
\end{equation}
where we have defined $r\equiv r\left(  x\right)  $ and%
\begin{equation}
\left\{
\begin{array}
[c]{c}%
U_{1}\left(  r\right)  =\frac{6}{r^{2}}\left(  1-\frac{b\left(  r\right)  }%
{r}\right)  +\left[  \frac{3}{2r^{2}}b^{\prime}\left(  r\right)  -\frac
{3}{2r^{3}}b\left(  r\right)  \right]  \\
U_{2}\left(  r\right)  =\frac{6}{r^{2}}\left(  1-\frac{b\left(  r\right)  }%
{r}\right)  +\left[  \frac{1}{2r^{2}}b^{\prime}\left(  r\right)  +\frac
{3}{2r^{3}}b\left(  r\right)  \right]
\end{array}
\right.  .\label{potentials}%
\end{equation}
Note that the coordinate $x$ is appropriate only for Schwarzschild.
Nevertheless, we find convenient use the same variable even in the case of the
naked singularity. We choose%
\begin{equation}
b\left(  r\right)  =\left\{
\begin{array}
[c]{c}%
2MG\qquad\text{for Schwarzschild}\\
2\bar{M}G\qquad\text{for a naked singularity }\left(  \bar{M}=-\left\vert
M\right\vert \right)
\end{array}
\right.  .
\end{equation}
We find that%
\begin{equation}
b\left(  r_{t}\right)  =r_{t}%
\end{equation}
only for Schwarzschild. $r_{t}$ is termed the throat and $r\in\left[
r_{t},+\infty\right)  $. Of course, for the negative Schwarzschild mass,
$r\in\left(  0,+\infty\right)  $. The potentials of the Lichnerowicz operator
\ref{spin2} simplify into%
\begin{equation}
\left\{
\begin{array}
[c]{c}%
U_{1}\left(  r\right)  =m_{1}^{2}\left(  r\right)  =\left[  \frac{6}{r^{2}%
}\left(  1-\frac{2MG}{r}\right)  -\frac{3MG}{r^{3}}\right]  \\
\\
U_{2}\left(  r\right)  =m_{1}^{2}\left(  r\right)  =\left[  \frac{6}{r^{2}%
}\left(  1-\frac{2MG}{r}\right)  +\frac{3MG}{r^{3}}\right]
\end{array}
\right.  ,
\end{equation}
for the Schwarzschild case and%
\begin{equation}
\left\{
\begin{array}
[c]{c}%
\bar{U}_{1}\left(  r\right)  =\bar{m}_{1}^{2}\left(  r\right)  =\frac{6}%
{r^{2}}+\frac{15\bar{M}G}{r^{3}}\\
\\
\bar{U}_{2}\left(  r\right)  =\bar{m}_{1}^{2}\left(  r\right)  =\frac{6}%
{r^{2}}+\frac{9\bar{M}G}{r^{3}}%
\end{array}
\right.  .
\end{equation}
for the naked singularity. In the Schwarzschild case, we get%
\begin{equation}
\left\{
\begin{array}
[c]{c}%
m_{1}^{2}\left(  r\right)  \geq0\qquad\text{when }r\geq\frac{5MG}{2}\\
m_{1}^{2}\left(  r\right)  <0\qquad\text{when }2MG\leq r<\frac{5MG}{2}\\
\\
m_{2}^{2}\left(  r\right)  >0\text{ }\forall r\in\left[  2MG,+\infty\right)
\end{array}
\right.  .\label{negU}%
\end{equation}
The functions $U_{1}\left(  r\right)  $ and $U_{2}\left(  r\right)  $ play the
r\^{o}le of two r-dependent effective masses $m_{1}^{2}\left(  r\right)  $ and
$m_{2}^{2}\left(  r\right)  $, respectively. In order to use the WKB
approximation, we define two r-dependent radial wave numbers $k_{1}\left(
r,l,\omega_{1,nl}\right)  $ and $k_{2}\left(  r,l,\omega_{2,nl}\right)  $%
\begin{equation}
\left\{
\begin{array}
[c]{c}%
k_{1}^{2}\left(  r,l,\omega_{1,nl}\right)  =\omega_{1,nl}^{2}-\frac{l\left(
l+1\right)  }{r^{2}}-m_{1}^{2}\left(  r\right)  \\
\\
k_{2}^{2}\left(  r,l,\omega_{2,nl}\right)  =\omega_{2,nl}^{2}-\frac{l\left(
l+1\right)  }{r^{2}}-m_{2}^{2}\left(  r\right)
\end{array}
\right.  \label{rwn}%
\end{equation}
for $r\geq\frac{5MG}{2}.$ When $2MG\leq r<\frac{5MG}{2}$, $k_{1}^{2}\left(
r,l,\omega_{1,nl}\right)  $ becomes%
\begin{equation}
k_{1}^{2}\left(  r,l,\omega_{1,nl}\right)  =\omega_{1,nl}^{2}-\frac{l\left(
l+1\right)  }{r^{2}}+m_{1}^{2}\left(  r\right)  .\label{rwn1}%
\end{equation}

\section{One loop energy Regularization and Renormalization}

\label{p3}The total regularized one loop energy density for the graviton is
\begin{equation}
\rho(\varepsilon)=\rho_{1}(\varepsilon)+\rho_{2}(\varepsilon)\,,
\end{equation}
where the energy densities, $\rho_{i}(\varepsilon)$ (with $i=1,2$), are
defined as%
\[
\rho_{i}(\varepsilon)=\frac{1}{4\pi}\mu^{2\varepsilon}\int_{\sqrt{m_{i}%
^{2}(r)}}^{\infty}\,d\omega_{i}\,\frac{\omega_{i}^{2}}{\left[  \omega_{i}%
^{2}-m_{i}^{2}(r)\right]  ^{\varepsilon-1/2}}%
\]%
\begin{equation}
=-\frac{m_{i}^{4}(r)}{64\pi^{2}}\left[  \frac{1}{\varepsilon}+\ln\left(
\frac{\mu^{2}}{m_{i}^{2}\left(  r\right)  }\right)  +2\ln2-\frac{1}{2}\right]
.\label{energy}%
\end{equation}
The two r-dependent effective masses $m_{1}^{2}\left(  r\right)  $ and
$m_{2}^{2}\left(  r\right)  $ can be cast in the following form%
\begin{equation}
\left\{
\begin{array}
[c]{c}%
m_{1}^{2}\left(  r\right)  =m_{L}^{2}\left(  r\right)  +m_{1,S}^{2}\left(
r\right)  \\
\\
m_{2}^{2}\left(  r\right)  =m_{L}^{2}\left(  r\right)  +m_{2,S}^{2}\left(
r\right)
\end{array}
\right.  ,
\end{equation}
where%
\begin{equation}
m_{L}^{2}\left(  r\right)  =\frac{6}{r^{2}}\left(  1-\frac{b\left(  r\right)
}{r}\right)  \label{mL}%
\end{equation}
and%
\begin{equation}
\left\{
\begin{array}
[c]{c}%
m_{1,S}^{2}\left(  r\right)  =\left[  \frac{3}{2r^{2}}b^{\prime}\left(
r\right)  -\frac{3}{2r^{3}}b\left(  r\right)  \right]  \\
m_{2,S}^{2}\left(  r\right)  =\left[  \frac{1}{2r^{2}}b^{\prime}\left(
r\right)  +\frac{3}{2r^{3}}b\left(  r\right)  \right]
\end{array}
\right.  .\label{m12s}%
\end{equation}
Essentially for the problem we are investigating, the term containing
$m_{L}^{2}\left(  r\right)  $ is a long range term and will be discarded in
this analysis. The zeta function regularization method has been used to
determine the energy densities, $\rho_{i}$. It is interesting to note that
this method is identical to the subtraction procedure of the Casimir energy
computation, where the zero point energy in different backgrounds with the
same asymptotic properties is involved. In this context, the additional mass
parameter $\mu$ has been introduced to restore the correct dimension for the
regularized quantities. Note that this arbitrary mass scale appears in any
regularization scheme. Eq.$\left(  \ref{lambda1loop}\right)  $ for the energy
density becomes%
\begin{equation}
\frac{\Lambda_{c}}{8\pi G}=\rho_{1}(\varepsilon)+\rho_{2}(\varepsilon
)\,.\label{rho}%
\end{equation}
Taking into account Eq.$\left(  \ref{energy}\right)  $, Eq.$\left(
\ref{rho}\right)  $ yields the following relationship%
\begin{equation}
\hspace{-1.5cm}\frac{\Lambda_{c}}{8\pi G}=\sum_{i=1}^{2}\frac{m_{i}^{4}%
(r)}{64\pi^{2}}\left[  \frac{1}{\varepsilon}\,+\ln\left(  \left\vert
\frac{4\mu^{2}}{m_{i}^{2}(r)\sqrt{e}}\right\vert \right)  \right]
.\label{Lambdac}%
\end{equation}
Thus, the renormalization is performed via the absorption of the divergent
part into the re-definition of the bare classical constant $\Lambda_{c}$
\begin{equation}
\Lambda_{c}\rightarrow\Lambda_{0,c}+\Lambda^{div},
\end{equation}
where%
\begin{equation}
\Lambda^{div}=\frac{G}{32\pi\varepsilon}\left(  m_{1}^{4}\left(  r\right)
+m_{2}^{4}\left(  r\right)  \right)  .
\end{equation}
The remaining finite value for the cosmological constant reads%
\begin{equation}
\frac{\Lambda_{0,c}}{8\pi G}=\sum_{i=1}^{2}\frac{m_{i}^{4}(r)}{64\pi^{2}}%
\ln\left(  \left\vert \frac{4\mu^{2}}{m_{i}^{2}(r)\sqrt{e}}\right\vert
\right)  =\rho_{eff}^{TT}\left(  \mu,r\right)  .\label{lambda0}%
\end{equation}
The quantity in Eq.$\left(  \ref{lambda0}\right)  $ depends on the arbitrary
mass scale $\mu.$ It is appropriate to use the renormalization group equation
to eliminate such a dependence. To this aim, we impose that\cite{RGeq}%
\begin{equation}
\frac{1}{8\pi G}\mu\frac{\partial\Lambda_{0,c}\left(  \mu\right)  }%
{\partial\mu}=\mu\frac{d}{d\mu}\rho_{eff}^{TT}\left(  \mu,r\right)
.\label{rg}%
\end{equation}
Solving it we find that the renormalized constant $\Lambda_{0}^{TT}$ should be
treated as a running one in the sense that it varies provided that the scale
$\mu$ is changing
\begin{equation}
\Lambda_{0,c}\left(  \mu,r\right)  =\Lambda_{0,c}\left(  \mu_{0},r\right)
+\frac{G}{16\pi}\left(  m_{1}^{4}\left(  r\right)  +m_{2}^{4}\left(  r\right)
\right)  \ln\frac{\mu}{\mu_{0}}.\label{lambdamu}%
\end{equation}
Substituting Eq.$\left(  \ref{lambdamu}\right)  $ into Eq.$\left(
\ref{lambda0}\right)  $ we find%
\begin{equation}
\frac{\Lambda_{0,c}\left(  \mu_{0},r\right)  }{8\pi G}=\sum_{i=1}^{2}%
\frac{m_{i}^{4}(r)}{64\pi^{2}}\ln\left(  \left\vert \frac{4\mu_{0}^{2}}%
{m_{i}^{2}(r)\sqrt{e}}\right\vert \right)  .\label{Lambda0c}%
\end{equation}
Eq.$\left(  \ref{Lambda0c}\right)  $ is the expression we shall use to
evaluate both the geometries.

\subsection{The Schwarzschild metric}

The Schwarzschild background is simply described by the choice $b\left(
r\right)  =2MG$. In terms of the induced cosmological constant of Eq.$\left(
\ref{Lambda0c}\right)  $, we get%
\begin{equation}
\frac{\Lambda_{0,c}\left(  \mu_{0},r\right)  }{8\pi G}=\frac{1}{64\pi^{2}}%
\sum_{i=1}^{2}\left(  \frac{3MG}{r^{3}}\right)  ^{2}\ln\left(  \left\vert
\frac{4r^{3}\mu_{0}^{2}}{3MG\sqrt{e}}\right\vert \right)  ,
\end{equation}
where we have used the assumption that $m_{i,L}^{2}(r)$ can be neglected. We
know that an extremum appears, maximizing the induced cosmological constant
for%
\begin{equation}
\frac{3MG\sqrt{e}}{4r^{3}\mu_{0}^{2}}=\frac{1}{\sqrt{e}} \label{sol}%
\end{equation}
and leading to%
\begin{equation}
\frac{\Lambda_{0,c}\left(  \mu_{0},r\right)  }{8\pi G}=\frac{\mu_{0}^{4}%
}{4e^{2}\pi^{2}}%
\end{equation}
or%
\begin{equation}
\frac{\Lambda_{0,c}\left(  \mu_{0},r\right)  }{8\pi G}=\left(  \frac
{3MG}{r^{3}}\right)  ^{2}\frac{1}{64\pi^{2}}\qquad r\in\left[  r_{t},\frac
{5}{4}r_{t}\right]  .
\end{equation}
Therefore, it appears that there exists a bound for $\Lambda_{0,c}$%
\begin{equation}
\frac{9}{256\pi^{2}r_{t}^{4}}\leq\frac{\Lambda_{0,c}\left(  \mu_{0},r\right)
}{8\pi G}\leq\frac{225}{4096\pi^{2}r_{t}^{4}}%
\end{equation}

\subsection{The Naked Schwarzschild metric}

The energy densities of Eq.$\left(  \ref{energy}\right)  $ can be used also
for the negative Schwarzschild mass. The only change is in the range of
integration of the energy integral which can be extended to $\omega=0$. The
final result does not change, then in Eq.$\left(  \ref{Lambda0c}\right)  $, we
can substitute $M$ with $\bar{M}$. Although the equation formally maintains
the same expression, the throat is no more there. This means that a piece of
$m_{i,L}^{2}(r)$ cannot be neglected and the two effective masses become%
\begin{equation}
\left\{
\begin{array}
[c]{c}%
m_{1}^{2}\left(  r\right)  =\frac{6}{r^{2}}+\frac{15\bar{M}G}{r^{3}}\\
\\
m_{2}^{2}\left(  r\right)  =\frac{6}{r^{2}}+\frac{9\bar{M}G}{r^{3}}%
\end{array}
\right.  .
\end{equation}
Since the effective mass grows approaching the singularity, we approximate
them close to $r=0.$ Thus, we get%
\begin{equation}
\left\{
\begin{array}
[c]{c}%
m_{1}^{2}\left(  r\right)  =\simeq\frac{15\bar{M}G}{r^{3}}\\
\\
m_{2}^{2}\left(  r\right)  \simeq\frac{9\bar{M}G}{r^{3}}%
\end{array}
\right.  .
\end{equation}
In this case, Eq.$\left(  \ref{Lambda0c}\right)  $ yields%
\begin{equation}
\frac{\Lambda_{0,c}^{naked}\left(  \mu_{0},r\right)  }{8\pi G}=\frac{1}%
{64\pi^{2}}\left[  \left(  \frac{15\bar{M}G}{r^{3}}\right)  ^{2}\ln\left(
\left\vert \frac{4r^{3}\mu_{0}^{2}}{15\bar{M}G\sqrt{e}}\right\vert \right)
+\left(  \frac{9\bar{M}G}{r^{3}}\right)  ^{2}\ln\left(  \left\vert
\frac{4r^{3}\mu_{0}^{2}}{9\bar{M}G\sqrt{e}}\right\vert \right)  \right]
.\label{Lambda0cN}%
\end{equation}
In order to find an extremum, it is convenient to define the following
dimensionless quantity%
\begin{equation}
\frac{9\bar{M}G\sqrt{e}}{4r^{3}\mu_{0}^{2}}=x,
\end{equation}
then Eq.$\left(  \ref{Lambda0cN}\right)  $ becomes%
\begin{equation}
\frac{\Lambda_{0,c}^{naked}\left(  \mu_{0},r\right)  }{8\pi G}=-\frac{\mu
_{0}^{4}}{4e\pi^{2}}\left[  x^{2}\ln x+\frac{25}{9}x^{2}\ln\left(  \frac
{5x}{3}\right)  \right]  .
\end{equation}
We find a solution when%
\begin{equation}
\bar{x}=\frac{1}{\sqrt{e}}\left(  \frac{3}{5}\right)  ^{\frac{25}{34}}%
\simeq0.417\label{sol1}%
\end{equation}
corresponding to a value of%
\begin{equation}
\frac{\Lambda_{0,c}^{naked}\left(  \mu_{0},r\right)  }{8\pi G}=\frac{\mu
_{0}^{4}}{4e^{2}\pi^{2}}\frac{17}{75}5^{\left(  \frac{9}{17}\right)
}3^{\left(  \frac{8}{17}\right)  }\simeq0.328\frac{\mu_{0}^{4}}{4e^{2}\pi^{2}%
}=0.328\frac{\Lambda_{0,c}\left(  \mu_{0},r\right)  }{8\pi G}.
\end{equation}
This means that%
\begin{equation}
\frac{\Lambda_{0,c}^{naked}\left(  \mu_{0},r\right)  }{\Lambda_{0,c}\left(
\mu_{0},r\right)  }=0.328<1.\label{rap}%
\end{equation}
Finally, we spend few words on the trace part contribution, which essentially
confirms what has been found in Ref.\cite{Remo}. Repeating the same procedure
for the trace operator, we find for both values of the Schwarzschild mass
$\left(  \pm M\right)  $, that the only consistent value of finding extrema is
that of vanishing $M$. This happens because solutions $\left(  \ref{sol}%
,\ref{sol1}\right)  $ create a constraint on $M,r$ and $\mu_{0}$ which cannot
be simultaneously satisfied for the graviton and for the trace term.

\section{Summary and Conclusions}

\label{p4}In Ref.\cite{Remo}, we considered how to extract information on the
cosmological constant using the Wheeler-De Witt equation. In this paper, even
if we have applied the same formalism, we have looked at the reversed idea,
namely the induced cosmological constant represents a certain amount of energy
density which varies with the choice of the underlying background. It is quite
natural of thinking to an arrangement of the various induced constants in such
a way to have a classification system which looks like to a spectrum of
geometries. Note that this method, in principle can be extended beyond the
Schwarzschild sector, in such a way to include all the spherically symmetric
metrics. Extensions to modified theories of gravity have also been studied
only for positive Schwarzschild mass\cite{CG}. It is interesting also to note
that it is possible to use such method, represented by Eq.$\left(
\ref{WDW2}\right)  $, not only for the induced cosmological constant, but even
for electric or magnetic charges, simply by replacing%
\begin{equation}
\frac{1}{V}\frac{\int\mathcal{D}\left[  g_{ij}\right]  \Psi^{\ast}\left[
g_{ij}\right]  \int_{\Sigma}d^{3}x\hat{\Lambda}_{\Sigma}\Psi\left[
g_{ij}\right]  }{\int\mathcal{D}\left[  g_{ij}\right]  \Psi^{\ast}\left[
g_{ij}\right]  \Psi\left[  g_{ij}\right]  }=\frac{1}{V}\frac{\left\langle
\Psi\left\vert \int_{\Sigma}d^{3}x\hat{\Lambda}_{\Sigma}\right\vert
\Psi\right\rangle }{\left\langle \Psi|\Psi\right\rangle }=-\frac{\Lambda_{c}%
}{\kappa}.
\end{equation}%
\begin{equation}
\hat{\Lambda}_{\Sigma}\rightarrow\hat{Q}_{\Sigma}%
\end{equation}
and%
\begin{equation}
-\frac{\Lambda_{c}}{\kappa}\rightarrow-\frac{1}{2}\int_{\Sigma}d^{3}%
x\sqrt{^{3}g}\rho_{e}%
\end{equation}
leading to the following eigenvalue equation\cite{RemoCharge}%
\begin{equation}
\frac{\left\langle \Psi\left\vert \int_{\Sigma}d^{3}x\hat{Q}_{\Sigma
}\right\vert \Psi\right\rangle }{\left\langle \Psi|\Psi\right\rangle }%
=-\frac{1}{2}\int_{\Sigma}d^{3}x\sqrt{^{3}g}\rho_{e}.\label{Hel}%
\end{equation}
$\hat{Q}_{\Sigma}$ is the charge operator containing only the gravitational
field. Coming back to the result $\left(  \ref{rap}\right)  $, we can conclude
that the Schwarzschild naked singularity has a lower value of ZPE compared to
the positive Schwarzschild mass. This means that, even if the order of
magnitude is practically the same, the naked singularity is less favored with
respect to the Schwarzschild wormhole. A further progress could be the study
of the unstable sector in our formalism to better understand the behavior of
the naked Schwarzschild background. Indeed, we have studied the spectrum in a
W.K.B. approximation with the following condition $k_{i}^{2}\left(
r,l,\omega_{i}\right)  \geq0,$ $i=1,2$. Thus to complete the analysis, we need
to consider the possible existence of nonconformal unstable modes, like the
ones discovered in Refs.\cite{Instability}. If such an instability appears,
this does not mean that we have to reject the solution. In fact in
Ref.\cite{Remo1}, we have shown how to cure such a problem. In that context,
a\ model of \textquotedblleft space-time foam\textquotedblright\ has been
introduced in a large $N$ wormhole approach reproducing a correct decreasing
of the cosmological constant and simultaneously a stabilization of the system
under examination. It could be interesting in the context of the multi-gravity
to examine what happens for a large number of naked singularities.

\end{document}